\documentclass[twocolumn]{aastex631}

\usepackage{subfigure} 

\received{June 13, 2024}
\accepted{September 4, 2024}

\begin{document}

\title{How does the critical torus instability height vary with the solar cycle?}

\author[0000-0001-7927-9291]{Alexander W. James}
\affiliation{Mullard Space Science Laboratory, University College London, Holmbury St Mary, Dorking, Surrey, RH5 6NT, UK}

\author[0000-0002-0053-4876]{Lucie M. Green}
\affiliation{Mullard Space Science Laboratory, University College London, Holmbury St Mary, Dorking, Surrey, RH5 6NT, UK}

\author[0000-0003-3571-8728]{Graham Barnes}
\affiliation{NorthWest Research Associates, 3880 Mitchell Lane, Boulder, CO 80301, USA}


\author[0000-0002-2943-5978]{Lidia van Driel-Gesztelyi}
\affiliation{Mullard Space Science Laboratory, University College London, Holmbury St Mary, Dorking, Surrey, RH5 6NT, UK}
\affiliation{LESIA, Observatoire de Paris, Universit\'{e} PSL, CNRS, Sorbonne Universit\'{e}, Univ. Paris Diderot, Sorbonne Paris Cit\'{e}, 5 place Jules Janssen, 92195 Meudon, France}
\affiliation{Konkoly Observatory, Research Centre for Astronomy and Earth Sciences, Hungarian Academy of Sciences, Konkoly Thege \'{u}t 15-17., H-1121, Budapest,
Hungary}

\author[0000-0001-9922-8117]{David R. Williams}
\affiliation{European Space Agency (ESA), European Space Astronomy Centre (ESAC), Camino Bajo del Castillo, s/n, Villanueva de la Ca{\~n}ada, Madrid, 28692 Spain}



\begin{abstract}


The ideal magnetohydrodynamic torus instability can drive the eruption of coronal mass ejections. 
The critical threshold of magnetic field strength decay for the onset of the torus instability occurs at different heights in different solar active regions, and understanding this variation could therefore improve space weather prediction.
In this work, we aim to find out how the critical torus instability height evolves throughout the solar activity cycle.
We study a significant subset of HMI and MDI Space-Weather HMI Active Region Patches (SHARPs and SMARPs) from 1996--2023, totalling 21584 magnetograms from 4436 unique active region patches.
For each magnetogram, we compute the critical height averaged across the main polarity inversion line, the total unsigned magnetic flux and the separation between the positive and negative magnetic polarities.
We find the critical height in active regions varies with the solar cycle, with higher (lower) average critical heights observed around solar maximum (minimum).
We conclude this is because the critical height is proportional to the separation between opposite magnetic polarities, which in turn is proportional to the total magnetic flux in a region, and more magnetic regions with larger fluxes and larger sizes are observed at solar maximum.
This result is noteworthy because, despite the higher critical heights, more CMEs are observed around solar maximum than at solar minimum.
{}

\end{abstract}

\keywords{Solar magnetic fields(1503) --- Solar cycle(1487) --- Solar coronal mass ejections(310)}


\section{Introduction} \label{sec:introduction}

Coronal mass ejections (CMEs) are a significant component of space weather that affects the Earth and human activities in near-Earth space. These eruptions of plasma and magnetic field can interact with the Earth's magnetic field to induce damaging electrical currents and accelerate harmful bursts of energetic particles.
We are continuously developing capabilities to forecast space weather and the impacts it may cause. CMEs can take anywhere from 1--3 days to reach the Earth after they erupt, so to be able to produce a forecast with a longer lead time than that, we need to be able to predict CMEs before they occur. 
This will require an improved understanding of the mechanisms involved in CME initiation.

One theory is that CMEs are driven by the ideal magnetohydrodynamic torus instability \citep{kliem2006torus}. Magnetic flux ropes in the solar atmosphere carry electric currents and tend to expand radially via the hoop force, but the ambient magnetic field surrounding the flux rope contributes a stabilising tension force \citep{bateman1978instabilities,kliem2006torus}. However, if the strength of the restraining magnetic field drops off sufficiently rapidly with height, radial expansion of the flux rope will be unstable, leading to runaway expansion and the eruption of the flux rope as a CME \citep{kliem2006torus}. 
We can quantify the gradient of the magnetic field with height using a parameter known as the decay index, $n$, where
\begin{equation}
    n = - \frac{d \ln{B_{\mathrm{ext,p}}}}{d \ln{R}}.
    \label{eqn:decay_index}
\end{equation}
Here, $B_{\mathrm{ext,p}}$ is the poloidal component of the magnetic field external to a magnetic flux rope (the component that contributes the restraining tension force; where the poloidal direction is perpendicular to the axial (toroidal) direction of the flux rope and $R$ is distance in the solar radial direction.
For a symmetric torus with a major axis much larger than its minor axis, the critical value of the decay index at and above which the torus instability sets in is $n_{\mathrm{c}}=1.5$  \citep{bateman1978instabilities,kliem2006torus}, however observations and simulations suggest values that range from $1<n_{\mathrm{c}}<2$ \citep{torok2007simulations,fan2007onset,demoulin2010criteria,cheng2020initiation}.
We can define the height at which the critical value of the decay index occurs as the critical height, $h_{\mathrm{c}}$, and this will vary in time and space as the magnetic field evolves. Beneath the critical height is a torus-stable zone, and the region above the critical height is a torus unstable zone \citep{sun2022torus}.

\citet{james2022evolution} studied a sample of 42 active regions, quantifying how magnetic flux, the separation between opposite polarities, and the critical height in each region evolved over time and how these changes correlated with CME occurrence. 
Firstly, it was found that the CME rate was twice as high during periods when magnetic flux in the active regions was increasing than when it was decreasing, interpreted as resulting from the emergence of magnetic flux that leads to injection of free magnetic energy and increased complexity in the photospheric inversion lines. 
Secondly, during these times of increasing magnetic flux, 63\% more CMEs occurred per unit time when the critical height in the source active region was increasing rather than decreasing over a period of several hours before the eruption.
This seems like a surprising result because a rising critical height should make it harder for a flux rope to become torus unstable and erupt as a CME.

To help understand this observation, it is helpful to note that the critical height in bipolar magnetic regions is strongly correlated with the distance between the flux weighted centres of opposite magnetic polarities \citep{wang2017critical,baumgartner2018eruptive,james2022evolution}. Rising critical heights are therefore often observed when photospheric magnetic polarities move apart from each other, which can occur on a timescale of hours in the early emergence phase of active regions as an observational manifestation of the emergence of $\Omega\mathrm{-loops}$ through the photosphere. 
Since emerging magnetic flux can cause magnetic polarities to move apart, which in turn leads to higher critical heights, it can be difficult to separate the roles changing magnetic fluxes and critical heights play in causing eruptions.
Furthermore, CMEs were observed during phases of all combinations of increasing/decreasing magnetic flux and increasing/decreasing polarity separation.
Still, the results of \citet{james2022evolution} suggest the increase of magnetic flux is more important for creating the environment in which CMEs occur than changes in the critical height. 

In addition to studying how the critical height changes over time, it is also important to understand how the critical height compares to the height of an embedded flux rope. 
It has been shown that the critical height is typically lower in complex multipolar regions than in simple bipolar regions \citep{torok2007simulations}, and that the most magnetically complex active regions can produce many CMEs \citep[\textit{e.g.},][]{green2002prolific,gopalswamy2005extreme}. 
Active region properties show some variation over the solar cycle \citep{guo2010complexity}, and with more active regions on the Sun around the cycle maximum, the overall magnetic complexity increases. 
More CMEs occur around solar maximum that at solar minimum \citep{webb1994cycle,webb2012cmes}, with CME rates correlating closely with the sunspot number. The simplest explanation for this is that there are more sunspot groups present to produce CMEs, but it is also possible that the rate of CMEs produced per active region varies with the solar cycle.
The above points raise the question of whether the critical height also shows a temporal variation on the timescale of a solar cycle.

In this study, we use data from across three solar cycles to investigate changes in the critical height and how they compare with the magnetic fluxes and separations between opposite polarities in solar magnetic regions.
We outline the data used in this work in Section \ref{sec:data} and describe our methods in Section \ref{sec:methods}. We present our results in Section \ref{sec:results}, discuss our findings in Section \ref{sec:discussion}, and summarise our conclusions in Section \ref{sec:conclusions}.

\section{Data} \label{sec:data}

The Solar and Heliospheric Observatory (SOHO; \citealp{domingo1995soho}) was launched on 2 December 1995. Onboard is the Michelson Doppler Imager (MDI; \citealp{scherrer1995mdi}), which took measurements from 23 April 1996 until 27 October 2010 that enable the line-of-sight component of the photospheric magnetic field to be determined at an image cadence of 96 minutes and a spatial pixel size of $2''$. On 11 February 2010, the Solar Dynamics Observatory (SDO; \citealp{pesnell2012SDO}) was launched, and its Helioseismic and Magnetic Imager (HMI; \citealp{scherrer2012hmi}) instrument produces 3D vector magnetograms at a cadence of 12 minutes with a pixel size of $0.5''$. We use magnetograms from the SHARP (Spaceweather HMI Active Region Patch; \citealp{bobra2014sharps}) and SMARP (Spaceweather MDI Active Region Patch; \citealp{bobra2021smarps}) datasets, which include cutouts of regions of significant photospheric magnetic flux that are tracked as they transit the solar disc. Each active region patch in the SHARP and SMARP datasets is assigned with a unique HARP or TARP (HMI/Tracked Active Region Patch) number, respectively, for identification, and they may contain zero or more active regions as designated by NOAA (the National Oceanic and Atmospheric Administration of the United States). 

In our work, we use HARPs and TARPs from the beginning of May 1996 until the end of October 2023, covering solar cycles 23 and 24, as well as the rising phase of solar cycle 25. We select regions that contain zero or one NOAA active region (patches with zero active regions are hereafter referred to as ephemeral regions), that were observed between $60^{\circ}$ east and west of central meridian (as observed by SOHO and SDO), and were significantly non-unipolar (\textit{i.e.}, at least $10\%$ of the total unsigned magnetic flux in the region was of the minority magnetic polarity, whether positive or negative). This non-unipolarity condition was enforced to ensure there would be clear polarity inversion lines in the regions of study, as these are essential to our method of computing the critical torus instability height (see Section \ref{sec:methods_hc}), and indeed are where pre-eruptive magnetic flux ropes necessarily form. 

\begin{figure}[!htb]
    \centering
    \includegraphics[width=0.45\textwidth]{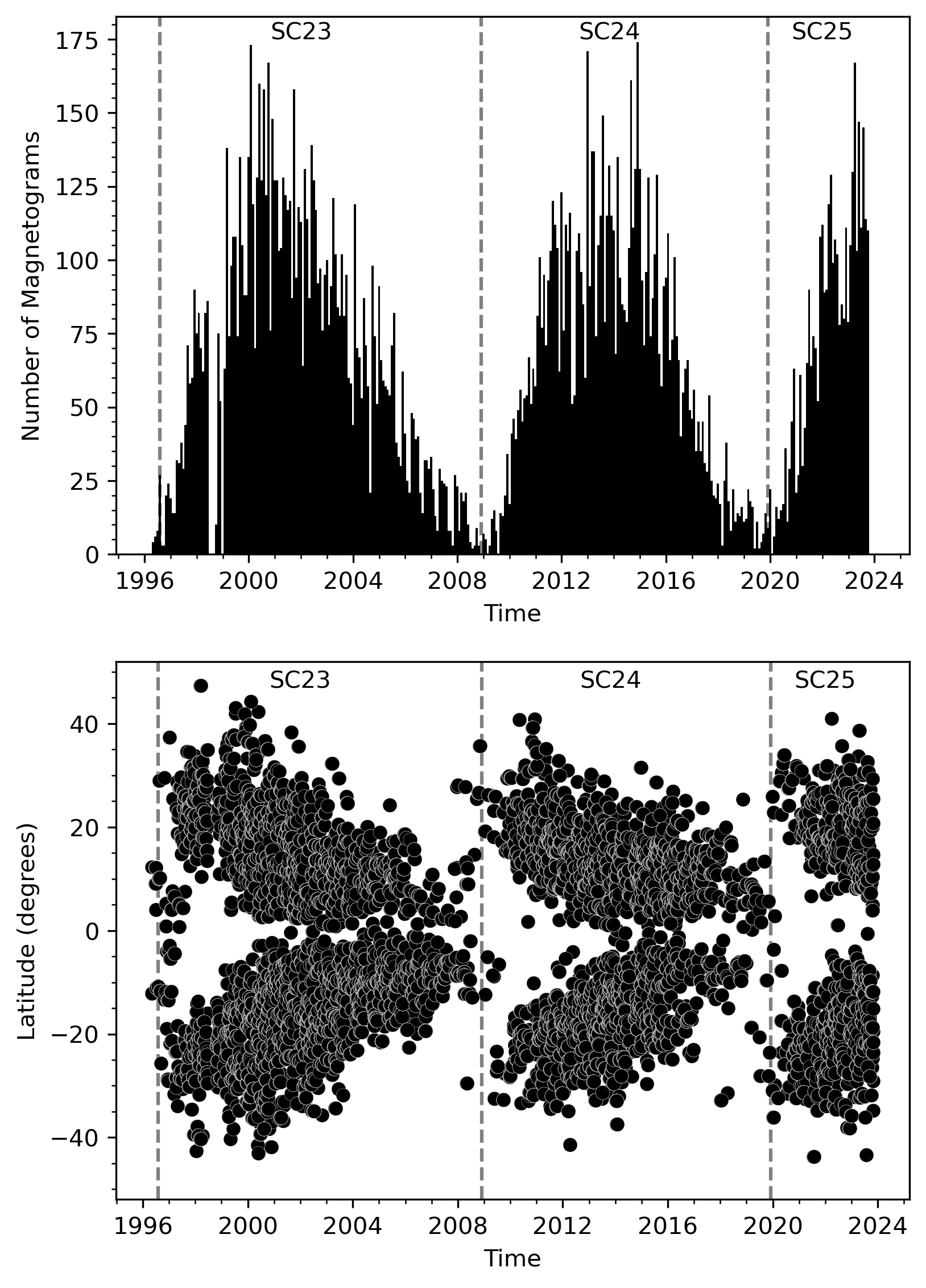}
    \caption{Top: Number of magnetograms in each month of our data set, which ranges from the beginning of May 1996 until the end of October 2023. During the overlap period of MDI and HMI observations from May--October 2010, we include only the number of HMI magnetograms to avoid double-counting. No MDI observations were made from 24 June 1998 until 19 October 1998 due to a loss of contact with the spacecraft. Bottom: Latitude of the flux-weighted centre of each magnetogram. The months of sunspot minima that separate solar cycles 23, 24 and 25 are indicated by vertical dashed lines.}
    \label{fig:mgms}
\end{figure}

To observe the evolution of the magnetic regions, we used magnetograms in a cylindrical equal area projection at an image cadence of 24 hours (taken at approximately 00:00 UT each day). 
In total, we used 21584 magnetograms from 4436 SMARPs and SHARPs (see Table \ref{tbl:num_arps_mgms} for a breakdown of the dataset across the ARP (Active Region Patch) types and the numbers of active regions and ephemeral regions).
The number of magnetograms used from each month is shown in the top panel of Figure \ref{fig:mgms}. Since we use magnetograms from TARPs and HARPs, it is no surprise that the monthly number of magnetograms in our sample varies with the solar activity cycle.
The ``butterfly diagram'' in the bottom panel of Figure \ref{fig:mgms} shows that the magnetograms in our study come from a wide range of active latitudes that evolves as expected throughout each sunspot cycle.

\begin{table}[!t]
\caption{Number of ARPs (top) and magnetograms (bottom) used in this study from the HMI and MDI datasets that contain one or zero NOAA active regions.}
\centering
\begin{tabular}{cccc}
\hline\hline
           \multicolumn{1}{c}{}  &  1 AR   &  0 ARs  &  Total \\
\hline
                     HARPs       &  1474   &   858   &  2332  \\
                     TARPs       &  1593   &   511   &  2104  \\
                     Total       &  3067   &  1369   &  4436  \\
\hline
\\
\hline\hline
                   \multicolumn{1}{c}{}     &  1 AR   &  0 ARs  &  Total \\
\hline
                    SHARP Magnetograms      &  8747   &  2758   &  11505 \\
                    SMARP Magnetograms      &  8678   &  1401   &  10079 \\
                                 Total      &  17425  &  4159   &  21584 \\
\hline
\end{tabular}
\label{tbl:num_arps_mgms}
\end{table}

We spatially resampled magnetograms from the SHARP dataset by a factor of 4 to match the $2''$ pixel size of the SMARP data. 
For SHARP magnetograms, we used magnetograms of the radial magnetic field component ($B_r$), however MDI magnetograms represent the line-of-sight field component. To approximate the radial field component in the SMARP magnetograms, we used the method of \citet{leka2017radial}, first performing a potential field extrapolation that uses the line-of-sight field component as the photospheric boundary condition to obtain a 3D vector magnetic field model, and then we take the $B_r$ component from the bottom of the extrapolation as a modelled magnetogram. As noted by \citet{leka2017radial}, this method generally performs better in active regions (where horizontal fields are significant) than a simple geometrical ``$\mu$-correction'' that assumes all of the observed magnetic field is radial. 

There is an approximately six-month overlap period in observations from when the SHARP series begins on 1 May 2010 and the SMARP series ends on 27 October 2010. 
During this time, we find 49 NOAA active regions that are in both the SMARP and SHARP datasets (and meet our selection criteria for non-unipolarity and disc longitude). 
Using the full MDI image cadence of 96 minutes (one in every eight HMI images), we find 3142 pairs of SHARP and SMARP magnetograms that were observed approximately co-temporally. 
In Section \ref{sec:methods_overlap}, we perform a comparison of the critical height values obtained from these SMARP and SHARP magnetograms.

\section{Methods} \label{sec:methods}

\subsection{Calculating the critical height} \label{sec:methods_hc}

To compute the decay index and therefore the critical height at which a flux rope would become unstable if present, we need the coronal magnetic field external to such a flux rope. We approximate the external magnetic field with the potential magnetic field, which we produce using the method of \citet{alissandrakis1981field} with $B_r$ magnetograms as the photospheric boundary condition ($B_r$ is obtained for SMARP data using the method described in Section \ref{sec:data}).
Without modelling flux ropes themselves, we cannot accurately determine the orientation of the poloidal magnetic field component, which is the necessary component in Equation \ref{eqn:decay_index}. Instead we approximate the poloidal component with the horizontal field component, as has been assumed in other studies \citep[e.g.][]{liu2008instabilites,zuccarello2015critical,wang2017critical,james2022evolution}.

We use a method similar to that performed in \citet{james2022evolution} to calculate the critical height. That is, polarity inversion lines (PILs) are identified by heavily smoothing the magnetograms with a $20 \times 20$ pixel spatial average and locating pixels where their neighbours are of the opposite magnetic polarity, the critical height is identified above each PIL pixel, and the mean critical height is calculated along the PIL. 
In some cases, the decay index exhibits a ``saddle''-shaped profile with height above the PIL \citep{guo2010confined,luo2022saddles}, in which there are multiple critical heights with a torus-stable zone between them. As in \citet{james2022evolution}, where multiple critical heights are found above one pixel, we select the lowest because these tend to occur at comparable heights to those calculated in regions where there is no saddle (typically ${\sim}40\ \mathrm{Mm}$). Furthermore, although saddle-shaped decay index profiles with multiple critical heights and an intermediate stable zone can cause failed or two-step eruptions \citep[e.g.][]{gosain2016interrupted,liu2018failed}, CMEs can still occur from such regions, particularly when the decay index at the minimum of the saddle is not too small \citep{wang2017critical} and when a large enough Lorentz force is applied to give the erupting structure enough momentum to traverse the torus-stable region \citep{liu2018failed}. Magnetic reconnection may also play a role in continuing to accelerate an erupting structure after an initial torus instability subsides.

\begin{figure}[!t]
    \centering
    \includegraphics[width=0.47\textwidth]{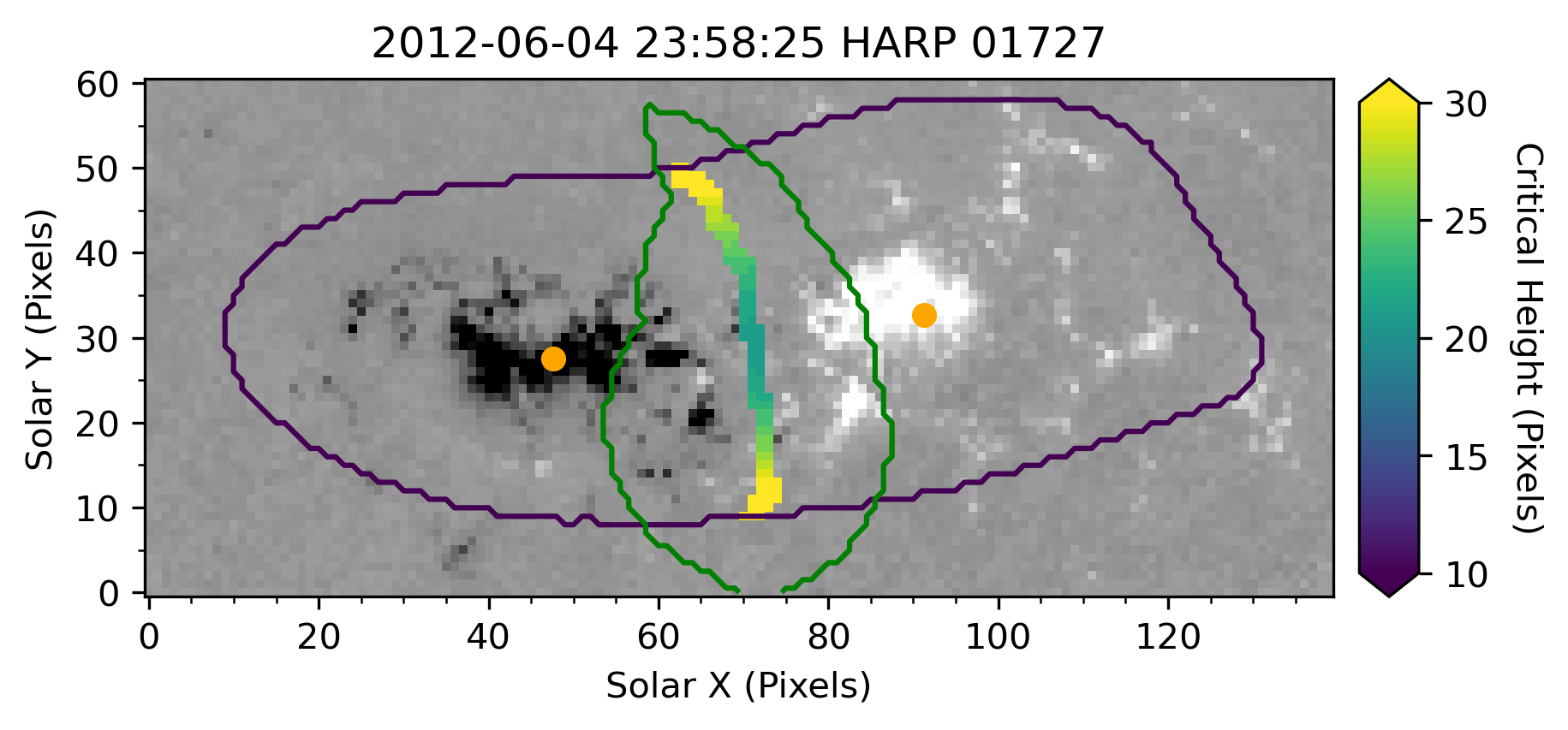}
    \caption{HARP 1727, containing NOAA AR 11497. ``Weak field'' bitmap pixels are outlined with the purple contour. The green contour outlines a region identified as being between strong opposite polarities. PIL pixels within both of these contours are identified and coloured by the local critical height. Orange circles show the flux-weighted centroids of positive and negative magnetic flux.}
    \label{fig:pil_hc}
\end{figure}

Whilst \citet{james2022evolution} manually defined sections of the detected PILs that were relevant, we implement an automated method due to the large number events in our sample that uses the field strength pixel bitmaps that are included with the SHARP and SMARP datasets. We select PIL pixels that are within the weak field bitmaps, but also within regions where dilated positive and negative strong field bitmaps overlap with each other (similarly to how the $R$ parameter is computed by \citealt{schrijver2007R} to quantify sharp gradients across inversion lines), resulting in masked PIL pixels that lie between strong concentrations of opposite polarity magnetic fluxes (see Figure \ref{fig:pil_hc}).  
Sometimes this method identifies PIL pixels that are further from the core of the active regions than we would like, and particularly in multipolar active regions, there is sometimes more than one distinct PIL identified, meaning the critical heights from each PIL are averaged into a single critical height for the magnetogram. However, the large number of magnetograms we are using should help to reduce the impact of outliers where the PILs are not identified as clearly.

\subsection{Quantifying magnetic flux and polarity separation}

To quantify the radial magnetic flux in each $B_r$ magnetogram, we first smooth the data using a $15 \times 15$ pixel spatial average. Then, we create a mask of pixels where the smoothed values are greater than a threshold magnitude of $100\, \mathrm{G}$. We apply this pixel mask to the unsmoothed magnetograms and integrate the positive and negative flux densities, thereby excluding weak field pixels where the signal-to-noise ratio is poor. Finally, we calculate the unsigned magnetic flux as the mean of the absolute values of positive and negative magnetic flux to give a single magnetic flux value for each magnetogram.

We quantify the characteristic separation between positive and negative magnetic polarities in each magnetogram using the same $15 \times 15$ pixel spatially smoothed magnetograms as are used to calculate the magnetic flux. Using only pixels where the magnitude of smoothed flux density is greater than $100\, \mathrm{G}$, we compute the flux-weighted centres of positive and negative magnetic flux. Examples of these two centroid coordinates are shown with orange circles in Figure \ref{fig:pil_hc}, and we use the distance between the pair of centroids as our measure of the separation between opposite magnetic polarities in each magnetogram.

\subsection{Comparison of critical heights from SMARP and SHARP extrapolations} \label{sec:methods_overlap}

Before we can make conclusions using observations from MDI and HMI, we need to understand any systematic differences between the datasets.
The magnetograms from the two instruments are produced using observations across different spectral lines, with MDI observing in a tunable band around the \ion{Ni}{1} 6768 \AA{} absorption line and HMI similarly around the \ion{Fe}{1} 6173 \AA{} absorption line.
Furthermore, unlike the line-of-sight MDI magnetograms, the HMI vector magnetograms are produced using the Very Fast Inversion of the Stokes Vector \citep{borrero2011VFISV}, and as outlined in Section \ref{sec:data}, HMI has a better spatial resolution than MDI.
Therefore, even when the two telescopes observe the same region of the Sun, there are likely to be some differences in the inferred magnetic fields.
Here, we calibrate the critical height values we obtain from MDI and HMI observations of the same active regions.

\begin{figure}[!t]
    \centering
    \includegraphics[width=0.47\textwidth]{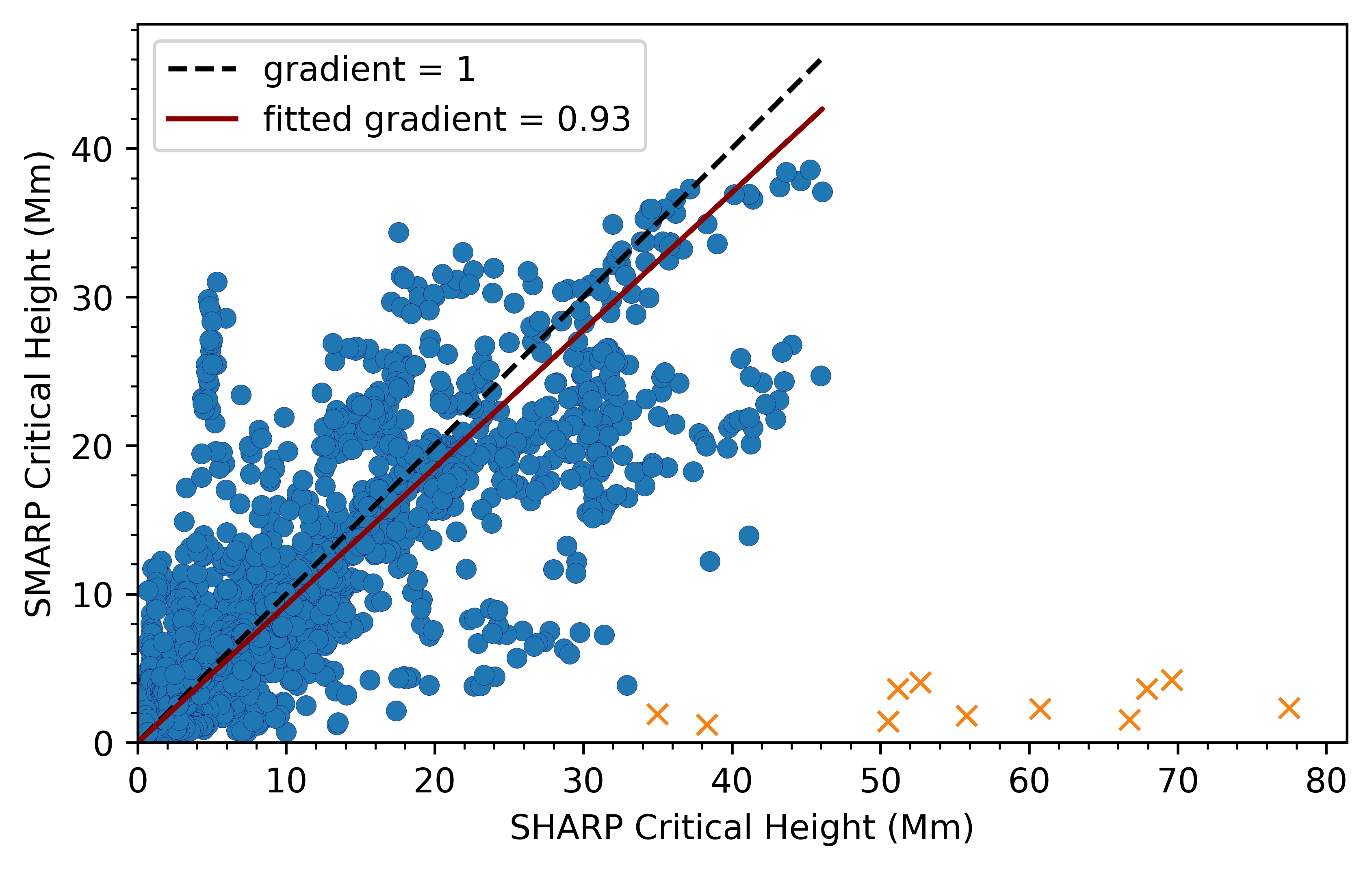}
    \caption{Mean critical heights calculated from 3142 SHARP and SMARP magnetogram pairs that were observed approximately cotemporally between 1 May and 28 October 2010, represented by blue circles. Outliers where the SHARP critical heights are much greater than the SMARP critical heights are depicted as orange crosses and excluded. The dashed black line illustrates a gradient of one for illustration, whereas the red line corresponds to a linear fit to the data (using orthogonal distance regression and constrained to pass through the origin) with a gradient of 0.93. The Pearson's correlation coefficient is $r=0.864$.}
    \label{fig:overlap_mdi_hmi}
\end{figure}

As introduced in Section \ref{sec:data}, we find 3142 pairs of SHARP and SMARP magnetograms that contain the same NOAA active region observed almost cotemporally. 
Even when observing the same regions at the same time, the fields-of-view of the SHARPs and SMARPs are generally different, so we crop and use only the common area that is contained within both magnetograms.
To account for the different spatial resolutions, we spatially resample the HMI magnetograms by a factor of 4 to match the MDI resolution.
Then, using the method described in Section \ref{sec:methods_hc}, we extrapolate potential coronal magnetic fields from the SHARP and SMARP magnetograms and calculate the mean critical height above photospheric PILs. 
For each SHARP-SMARP magnetogram pair, we identified PIL pixels using the mask made from the SHARP data to ensure the same pixels were used to calculate the critical height in both datasets.
We obtain critical heights from the 3142 pairs of SHARP and SMARP magnetograms during the period of overlapping observations, and these are presented in Figure \ref{fig:overlap_mdi_hmi}.

A number of outliers where the critical heights calculated from the SHARP magnetograms are significantly larger than from the SMARPS can be seen in Figure \ref{fig:overlap_mdi_hmi} (marked with orange `X' symbols). 
Specifically, we exclude 11 points where the SHARP critical height is greater than $34\, \textrm{Mm}$ but the critical height from a SMARP magnetogram at the same time is $< 5\, \textrm{Mm}$. All of these datapoints come from early in the observed lifetime of HARP 00224, when it contained the weak and dispersed remnants of a decayed active region before NOAA 11119 emerged into it. The weak field bitmap used to search for PIL pixels between opposite polarities (described in Section \ref{sec:methods_hc}) only covered the negative polarity, so very few PIL pixels were identified. Despite the fact that we compute critical heights using the same PIL pixels in both the SHARP and SMARP magnetograms, the critical height values obtained by averaging across so few PIL pixels are unreliable. We exclude the values obtained from these 11 magnetograms from our study.

We perform an orthogonal distance regression fit to the SMARP and SHARP critical height pairs to see how they compare. Constraining the fit to pass through the origin, we find a gradient of 0.93 with a Pearson's correlation coefficient of $r=0.838$. This suggests that the critical heights obtained from SMARP magnetograms are, on average, slightly lower than from the corresponding SHARP magnetograms. 
However, whilst there are magnetograms where the SHARP critical height is higher than the paired SMARP value, there are also many magnetograms where the opposite is true. We take the ratio of the SHARP-to-SMARP critical height for each magnetogram pair, and find the median ratio across the dataset to be 0.99. This suggests there are roughly as many magnetogram pairs where the SHARP critical height is higher as there are pairs where the SMARP critical height is higher, and therefore no systematic scaling is required for either the HMI or MDI critical heights.
Finally, we perform the Student's t-test and the Wilcoxon signed rank test on the HMI and MDI critical heights and find p-values of $0.40$ and $0.93$, respectively. Both of these are significantly greater than $0.05$, so we cannot reject the null hypothesis. In other words, there is no evidence to suggest the HMI and MDI critical heights are meaningfully different from one another.

Whereas above we cropped the commonly-observed field-of-view from from each pair of SHARP and SMARP magnetograms and averaged their critical heights using the same mask of PIL pixels made using the SHARP data and bitmap, we repeat the analysis whilst treating the SHARP and SMARP magnetograms independently as an additional test of how well the results obtained from each agree. Whilst the spatial resolution of the SHARP magnetograms is still downsampled by a factor of 4 (matching that of the SMARP magnetograms) in order to save computational time, this time we compute critical heights with different pixel masks in the SHARP and SMARP magnetograms. The full SHARP and SMARP fields-of-view are considered and PIL pixels are identified independently for each dataset using their respective bitmaps and $B_{r}$. 
The independently-computed SHARP and SMARP critical heights are still in very close agreement, albeit slightly less so than we found using the previous method.

We now only find and exclude 3 magnetograms that fit the definition for outliers we used above (\textit{i.e.} where the SHARP critical height is greater than $34\, \textrm{Mm}$ but the critical height from a SMARP magnetogram at the same time is $< 5\, \textrm{Mm}$). These magnetograms still correspond to NOAA 11119. 
The Pearson's correlation coefficient of the SHARP and SMARP critical heights using this method is $r=0.762$ (compared to the previous $r=0.838$) and an orthogonal distance regression fit results in a gradient of 0.87 (compared to the previous 0.93), suggesting the average critical height from the SHARP magnetograms is greater than from the SMARP magnetograms by a little bit more than was seen by using the other method (\textit{i.e.} slightly further away from a ratio of 1.0). 
We again calculate the ratio of the critical heights from each SHARP-SMARP pair and take the median ratio across the dataset, resulting in a ratio of 1.05. This is slightly larger than the ratio of 0.99 found using the previous method, but is still very close to unity, suggesting no significant systematic difference between the HMI and MDI critical heights.

\section{Results} \label{sec:results}

We investigate how the magnetic flux, polarity separation and critical height vary throughout the SMARP and SHARP datasets.

\subsection{Active region flux, polarity separation, and critical height} \label{sec:results_flux_sep_hc}

\begin{figure}[!t]
    \centering
    \includegraphics[width=0.45\textwidth]{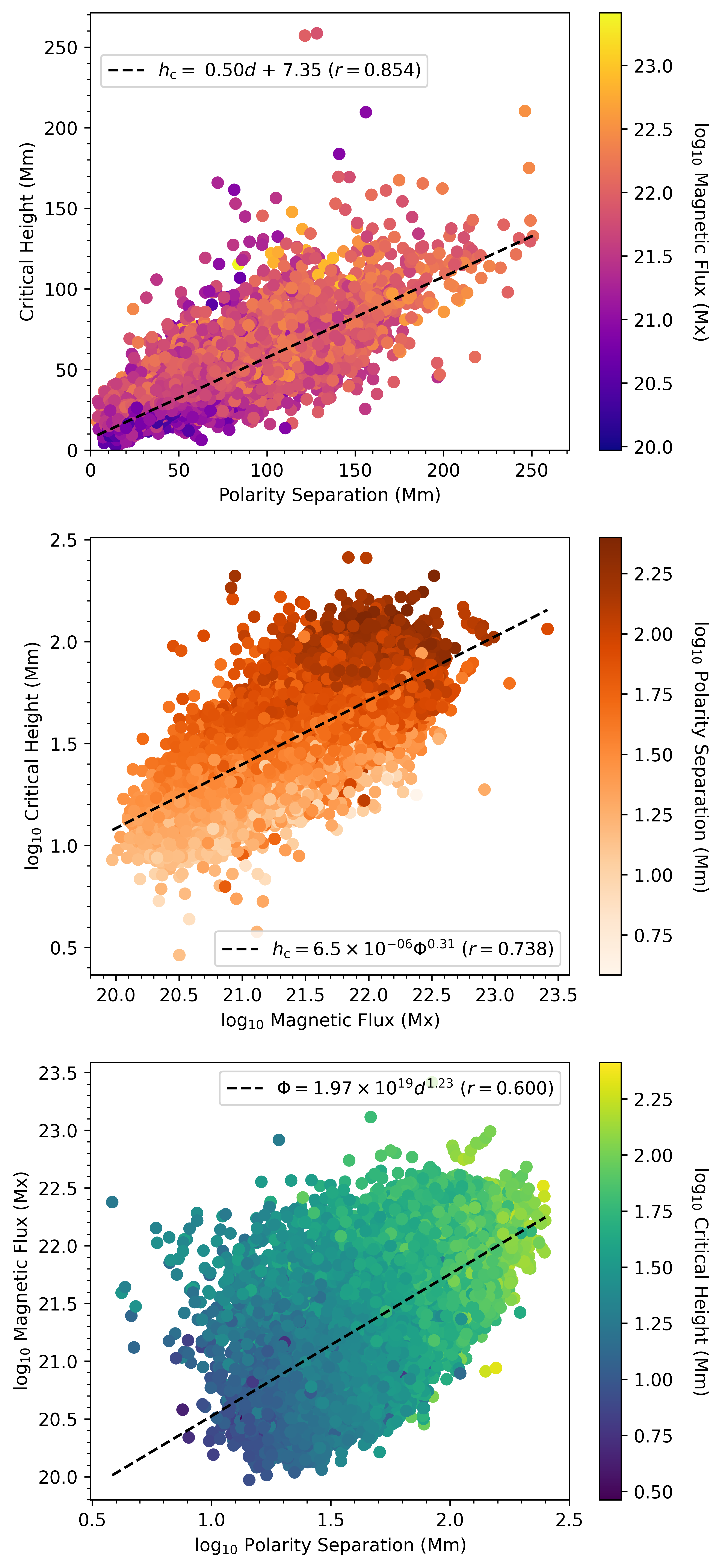}
    \caption{Top: critical height ($h_{\mathrm{c}}$) vs polarity separation ($d$), coloured by the logarithm of the unsigned magnetic flux ($\Phi$). Middle: critical height vs unsigned magnetic flux (log-log scale), coloured by the logarithm of the polarity separation. Bottom: magnetic flux vs polarity separation (log-log scale), coloured by the logarithm of the critical height. Each point represents one of 21584 magnetograms. Linear fits to the data are indicated by black dashed lines, and the equations of the lines are given in the legends. The Pearson's correlation coefficients for the quantities on the $x-y$ axes of each panel are $r=0.854$ in the top panel, $r=0.7388$ in the middle panel, and $r=0.600$ in the bottom panel.}
    \label{fig:correlation_plots}
\end{figure}

In the top panel of Figure \ref{fig:correlation_plots}, we plot the calculated critical height ($h_\mathrm{c}$) against the measured separation between opposite polarities ($d$) for each of the 21584 magnetograms. The critical height and polarity separation are strongly correlated, with a Pearson correlation coefficient ($r$) of 0.854, and we perform a linear fit to the data, resulting in the relationship $h_\mathrm{c} = 0.5d + 7.35$. 

The colour of each datapoint represents the unsigned magnetic flux calculated in that magnetogram. 
The smallest (largest) magnetic fluxes are generally seen in the magnetograms with the smallest (largest) polarity separations and lowest (highest) critical heights.
Specifically, many of the weaker magnetic fluxes (${<}10^{21}\, \textrm{Mx}$) are measured in magnetograms with small polarity separations (${<}80\, \textrm{Mm}$) and small critical heights (${<}40\, \textrm{Mm}$).
However, a small population of magnetograms with polarity separations $50<d<170\, \textrm{Mm}$ and critical heights $h_{\mathrm{c}}>100\, \textrm{Mm}$ are seen to have abnormally small magnetic fluxes $<10^{21}\, \textrm{Mx}$, and there are also two magnetograms with exceptionally high critical heights $>250\, \textrm{Mm}$ for their polarity separations ($\approx125\, \textrm{Mm}$).
These magnetograms all contain magnetic flux concentrations that are very close to quiet Sun conditions, captured either early in the ARP's life before significant flux emergence has taken place, or after the region's magnetic flux has decayed.

In the middle panel of Figure \ref{fig:correlation_plots}, we show the critical height against the unsigned magnetic flux in a log-log scale for each of the 21584 magnetograms, with colours corresponding to the polarity separation. The Pearson correlation coefficient between these two quantities is $r=0.738$, and a linear fit to the data gives $h_{\mathrm{c}} = 6.5 \times 10^{-6}\, \Phi^{0.31}$, where $\Phi$ is measured in $\textrm{Mx}$ and $h_{\mathrm{c}}$ is in $\textrm{Mm}$.
The magnetograms with the largest polarity separations are generally seen in magnetograms with the largest magnetic fluxes and that have the largest critical heights.
There are a few outliers with polarity separations $>10^{2}\, \textrm{Mm}$ and critical heights $<10^{1.4}\, \textrm{Mm}$.

In the bottom panel of Figure \ref{fig:correlation_plots}, we plot the magnetic flux (measured in $\mathrm{Mx}$) against the polarity separation (in $\mathrm{Mm}$) from all 21584 magnetogtams using a log-log scale. These parameters have a Pearson correlation coefficient of $r=0.600$, and a linear fit to the data gives the relationship $\Phi = 1.97 \times 10^{19}\, \mathrm{Mx}\, d^{1.23}$, where $d$ is measured in $\mathrm{Mm}$ and $\Phi$ is in $\textrm{Mx}$.
For direct comparison with the relationship obtained by \citet{wang1989bipolar} (presented in Section \ref{sec:discussion}), we also perform the fit with $d$ measured in degrees, finding the relationship $\Phi = 4.23 \times 10^{20}\, \mathrm{Mx}\, d^{1.23}$.

We also repeat the linear fitting of unsigned flux against polarity separation after separating the magnetograms that contain a NOAA active region from those which do not (ephemeral regions). 
With separations, $d$, in $\textrm{Mm}$, we find $\Phi = 3.27 \times 10^{19}\, \mathrm{Mx}\, d^{1.14}$ for NOAA active regions ($r=0.566$) and $\Phi = 7.42 \times 10^{18}\, \mathrm{Mx}\, d^{1.31}$ in ephemeral regions ($r=0.772$).
When $d$ is measured in degrees, these relationships are $\Phi = 5.64 \times 10^{20}\, \mathrm{Mx}\, d^{1.14}$ for NOAA active regions and $\Phi = 1.95 \times 10^{20}\, \mathrm{Mx}\, d^{1.31}$ in ephemeral regions.

\subsection{Temporal evolution of critical height, polarity separation, and magnetic flux} \label{sec:results_solar_cycle}

\begin{figure*}[!pt]
    \centering
    \begin{subfigure}
        \centering
        \includegraphics[width=0.44\textwidth]{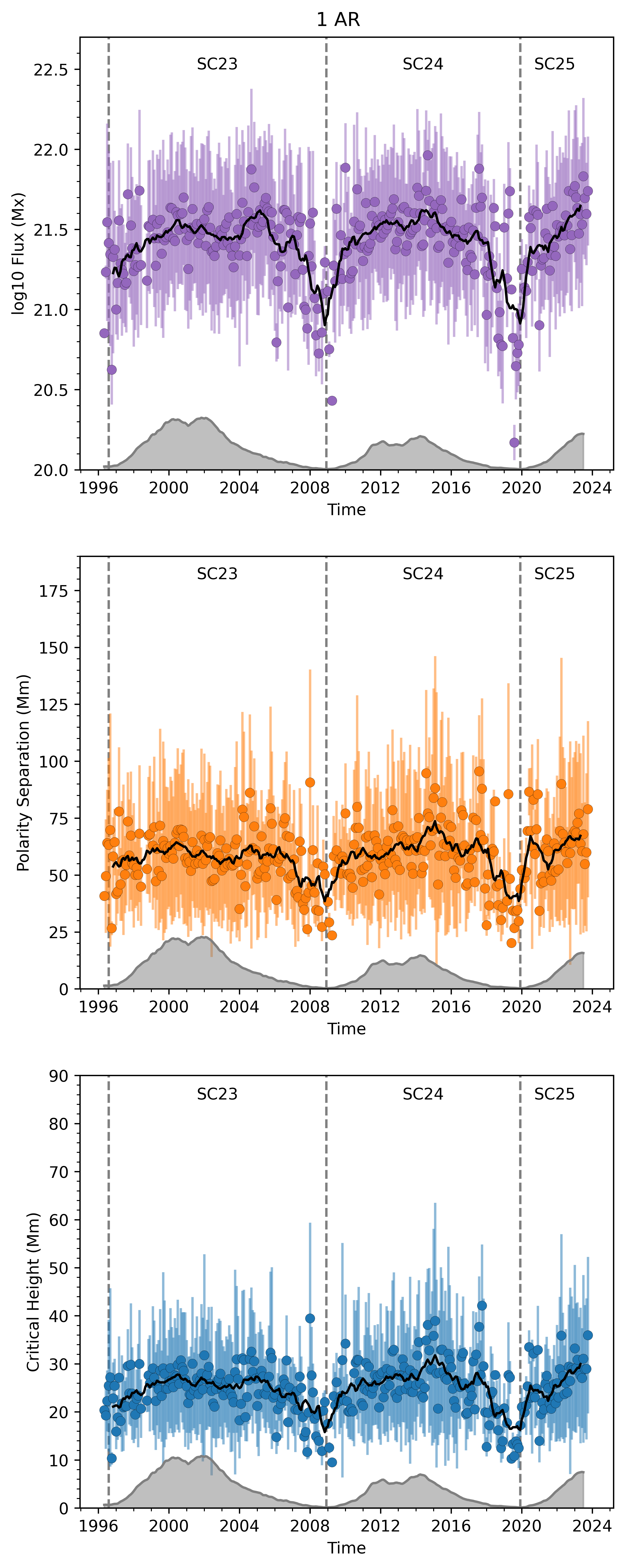}
    \end{subfigure}%
    ~ 
    \begin{subfigure}
        \centering
        \includegraphics[width=0.44\textwidth]{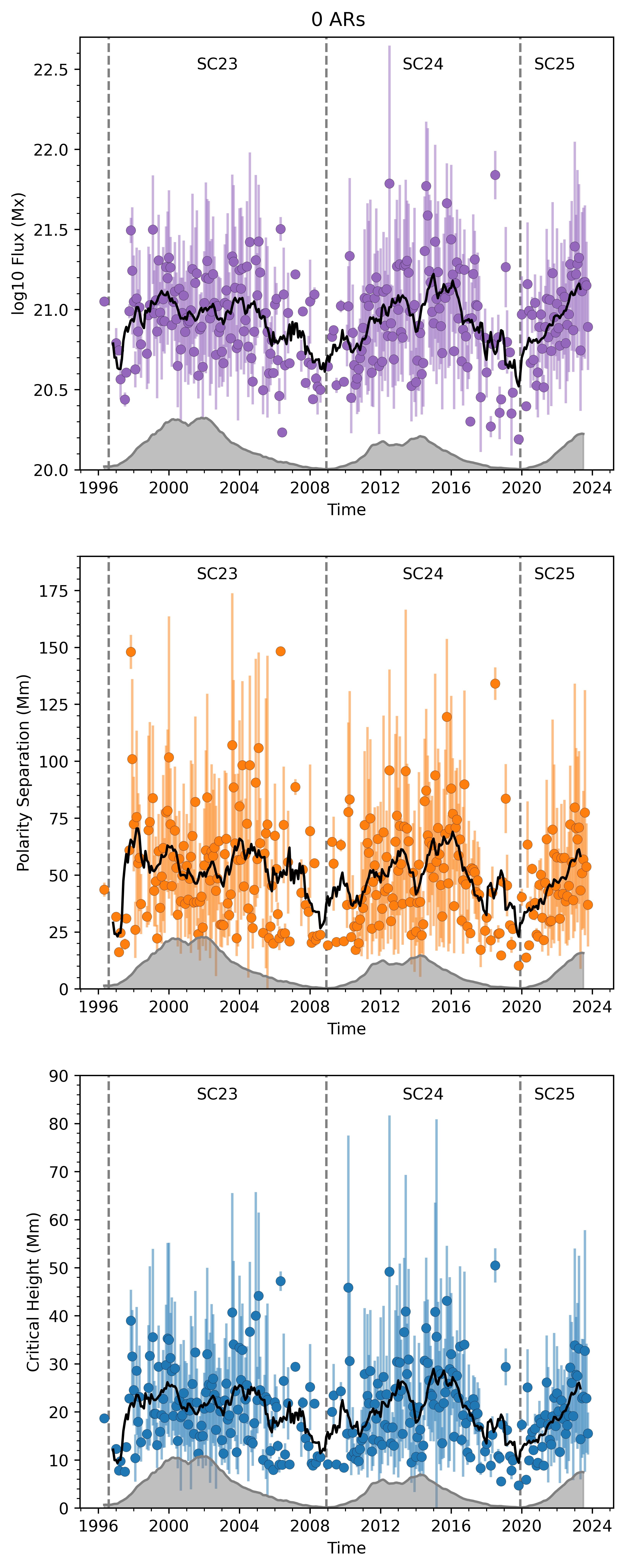}
    \end{subfigure}
    \caption{The left column shows results from ARPs that contain one NOAA active region and the right column represents HARPs that contain ephemeral regions. Mean unsigned magnetic fluxes (top), mean separations between opposite magnetic polarities (middle), and mean critical heights (bottom) of all magnetograms in each month of our dataset are marked with filled circles, and the standard deviation observed in each month are indicated with vertical bars. The 12-month moving means are represented by black curves. The 13-month smoothed monthly sunspot number is represented by the grey shaded area along the bottom of each panel, and the sunspot minima of August 1996, December 2008 and December 2019 are indicated by by vertical dashed lines to show the start times of solar cycles 23, 24 and 25.}
    \label{fig:cycle_variations}
\end{figure*}

In Figure \ref{fig:cycle_variations}, we show the mean unsigned flux (top), the mean separation between the flux-weighted centres of positive and negative magnetic flux (middle), and the mean critical height (bottom) of all the magnetograms in each month of our data sample containing a single NOAA active region (left column) and ephemeral regions (right).
The standard deviations of the values across all magnetograms in each month are presented as errorbars to indicate the observed ranges in each parameter.
Though the standard deviations, $\sigma$, are often large, particularly around solar maximum, the standard error ($\sigma/\sqrt{N}$) in the monthly means is generally very low due to the large number of magnetograms sampled, $N$. This suggests the monthly mean values are well-constrained.
For comparison, we also show the 13-month smoothed sunspot number (issued by the Solar Influence Data Analysis Center) in each panel of Figure \ref{fig:cycle_variations}, and we indicate the times of sunspot minima which demarcate solar cycles 23, 24 and 25. These took place in August 1996, December 2008 and December 2019.

For the NOAA active regions, the lowest monthly means of unsigned magnetic flux, polarity separation, and critical height are seen around 1996, 2008 and 2019 --- around the same times as the sunspot minima --- with typical values $< 10^{21}\, \mathrm{Mx}$, $25\textrm{--}40\, \mathrm{Mm}$, and $10\textrm{--}20\, \mathrm{Mm}$, respectively. 
There is relatively little range in the values observed in any given month around these times of lower solar activity.

Periods of consistently higher monthly mean values occur from roughly 1998--2007, 2010--2017, and from 2021 until the end of our dataset in October 2023, with unsigned fluxes ${>} 10^{21}\, \mathrm{Mx}$, polarity separations of $50\textrm{--}80\, \mathrm{Mm}$, and critical heights of $20\textrm{--}35\, \mathrm{Mm}$. These periods span the years around the times of maximum observed sunspot numbers in solar cycles 23 and 24, and appear to fit with the increasing numbers of sunspots in solar cycle 25 that are currently predicted to peak in 2024 or 2025 \citep{bhowmik2018prediction,mcintosh2023terminator,noaa2023update}.
Although the monthly means are higher during these periods of higher solar activity, the spreads seen each month show that values from individual magnetograms can still be almost as low as those seen around solar minimum. 
However, significantly higher values of unsigned magnetic flux are also seen, with unsigned magnetic fluxes up to ${\sim} 10^{22}\, \mathrm{Mx}$, polarity separations up to ${\approx} 150\, \mathrm{Mm}$, and critical heights of up to ${\approx} 60\, \mathrm{Mm}$ in many months. This demonstrates the larger range of magnetic fluxes, polarity separations, and critical heights that are seen during periods of increased solar activity. 

The ephemeral regions also do not exhibit a wide range of values around solar minimum, with very few high values of magnetic flux, polarity separation and critical height around these times (typical mean monthly fluxes $\sim 10^{20}\, \mathrm{Mx}$, polarity separations $\approx 30\, \mathrm{Mm}$ and critical heights $\approx 15\, \mathrm{Mm}$). However, the quantities vary significantly from month to month around solar maximum, from some of the highest observed values (fluxes approaching $10^{22}\, \mathrm{Mx}$, polarity separations ${>}100\, \mathrm{Mm}$ and critical heights ${>} 40\, \mathrm{Mm}$) to lows comparable to those seen around solar minimum.

\section{Discussion} \label{sec:discussion}

\subsection{The correlation of magnetic flux, polarity separation, and critical height}

We study the relationships between the total unsigned magnetic flux ($\Phi$), the separation between the flux-weighted centres of positive and negative magnetic polarity concentrations ($d$), and the critical height $h_{\mathrm{c}}$ in 21584 magnetograms from May 1996 -- October 2023.

Firstly, we find that the critical height above PILs is positively correlated with the separation between positive and negative polarities, resulting in a relationship of $h_\mathrm{c} = 0.50d + 7.35$ and a strong Pearson's correlation coefficient of $r=0.854$. This is in good agreement with the findings of previous studies ($\frac{h_{\mathrm{c}}}{d}=0.54$; \citealp{wang2017critical}, $\frac{h_{\mathrm{c}}}{d}=0.4\pm0.1$; \citealp{baumgartner2018eruptive}, $\frac{h_{\mathrm{c}}}{d}=0.52\pm0.04$; \citealp{james2022evolution}). 

The outliers in the top panel of Figure \ref{fig:correlation_plots} that have small magnetic fluxes and relatively large critical heights (mentioned in Section \ref{sec:results_flux_sep_hc}) correspond to magnetograms that show approximately quiet Sun conditions. These may be captured either early in the region's life before significant flux emergence has taken place, or after the region's magnetic flux has decayed. In these quiet-Sun-like conditions, there are no strong, localised bipoles, so there are no strong field gradients (and therefore no critical heights) low down in the corona. Instead, we suggest the coronal magnetic field in the quiet Sun is generally dominated by large-scale global fields from distant bipoles, which decay over larger length scales and result in higher critical heights.

Secondly, the critical height also correlates closely with magnetic flux.
\citet{li2020fluxcritheight} identified a correlation between $h_{\mathrm{c}}$ and log$\Phi$, but in our dataset, we find a better fit when considering log($h_{\mathrm{c}}$) and log($\Phi$) (Pearson's correlation coefficient $r=0.738$). We fit the relation $h_{\mathrm{c}} = 6.5 \times 10^{-6}\, \Phi^{0.31}$, where $h_{\mathrm{c}}$ is in $\textrm{Mm}$ and $\Phi$ is in $\mathrm{Mx}$.

The outliers in the middle panel of Figure \ref{fig:correlation_plots} that have large polarity separations and relatively small critical heights (mentioned in Section \ref{sec:results_flux_sep_hc}) come from magnetograms that contain multiple magnetic bipoles or one strong bipole and another region of dispersed flux. 
This causes our method of calculating the separation between opposite magnetic polarities to perform poorly, as the positions of the two flux-weighted centres of positive and negative flux are averaged between the multiple bipoles and/or the regions of dispersed magnetic flux.
Furthermore, when more than one distinct PIL is identified in multipolar configurations, we average the critical heights from each PIL into a single value for the magnetogram, even though the critical heights can vary significantly from one PIL to another. 

Thirdly, we show that the logarithm of the separation between opposite magnetic polarities, $d$, correlates with the total magnetic flux in an observed region, $\Phi$, with a relationship of $\Phi \sim d^{1.23}$ and a Pearson's correlation coefficient of $r=0.600$ (in log-log space). This is in good agreement with the findings of previous studies ($\Phi \sim d^{1.3}$; \citealp{wang1989bipolar}, $\Phi \sim d^{1.15}$; \citealp{tian2003bipolar}). 

\citet{tian2003bipolar} studied only bipolar magnetic regions with two sunspots, and when we examine only the magnetograms from our dataset that contain a NOAA active region, we find $\Phi \sim d^{1.14}$ ($r=0.566$), which agrees very well with their result. 
Conversely, if we fit to only the magnetograms in our dataset that do not contain a NOAA active region (weaker, ephemeral regions), we find $\Phi \sim d^{1.31}$ ($r=0.772$). This is very close to the \citet{wang1989bipolar} result, although they considered a broad range of magnetic regions with and without sunspots, from strong regions $\sim 10^{22}\, \mathrm{Mx}$ down to weaker (likely ephemeral) regions $\sim 10^{20}\, \mathrm{Mx}$.

In summary, regions with larger magnetic fluxes tend to have larger separations between their opposite magnetic polarities, and regions with larger polarity separations tend to have higher critical heights. 
Therefore, it follows that, since the magnetic flux in our studied magnetograms varies with the solar cycle, so too does the separation between opposite polarities and the critical height (seen clearly in Figure \ref{fig:cycle_variations}).

\subsection{Solar cycle variation of the critical height}

In active regions, the mean critical height for the onset of the torus instability varies with the solar cycle, as does the unsigned magnetic flux and the separation between opposite polarities. Minimum smoothed sunspot numbers were observed in 1996, 2008 and 2019, around which times we see the smallest mean monthly magnetic fluxes ($< 10^{21}\, \mathrm{Mx}$), the shortest polarity separations ($30\textrm{--}50\, \mathrm{Mm}$) and the lowest monthly critical heights ($15\textrm{--}20\, \mathrm{Mm}$).
Maximum smoothed sunspot numbers were observed in late 2001 and 2014 (with another maximum expected around 2025), \textit{i.e.}, in the middles of the extended periods when we see consistently higher mean monthly unsigned fluxes (${>} 10^{21}\, \mathrm{Mx}$), polarity separations ($60\textrm{--}100\, \mathrm{Mm}$) and critical heights ($30\textrm{--}45\, \mathrm{Mm}$).

The solar cycle variation of these quantities is less clear in ephemeral regions. The largest monthly means of magnetic flux, polarity separation, and critical height are observed around solar minimum, but there are also many months around solar maximum with very low mean quantities. In other words, there is greater variance in the monthly means around solar maximum, but comparably low values are still observed throughout each solar cycle.

There are naturally fewer magnetograms in our dataset around solar minimum, and this will enable extreme values to more strongly affect the mean and the spread of values seen in the magnetograms each month.
However, at least for active regions, it is still interesting that the fluxes, polarity separations and critical heights at solar minimum are generally amongst the lowest values observed, with very few large fluxes, separations, or critical heights identified (as evidenced by the small spreads around these times in Figure \ref{fig:cycle_variations}).
We note that the few regions with strong magnetic fluxes (${>}10^{22}\, \textrm{Mx}$) seen around periods of solar minimum are all from latitudes lower than $12^{\circ}$, and therefore likely belong to the waning solar cycle, whereas regions with with weaker magnetic fluxes (${<}5\times10^{20}\, \textrm{Mx}$) are found from a wide range of latitudes up to $\approx 30^{\circ}$, which could represent the earliest emergences of the next cycle.

\begin{figure}[!htb]
    \centering
    \includegraphics[width=0.45\textwidth]{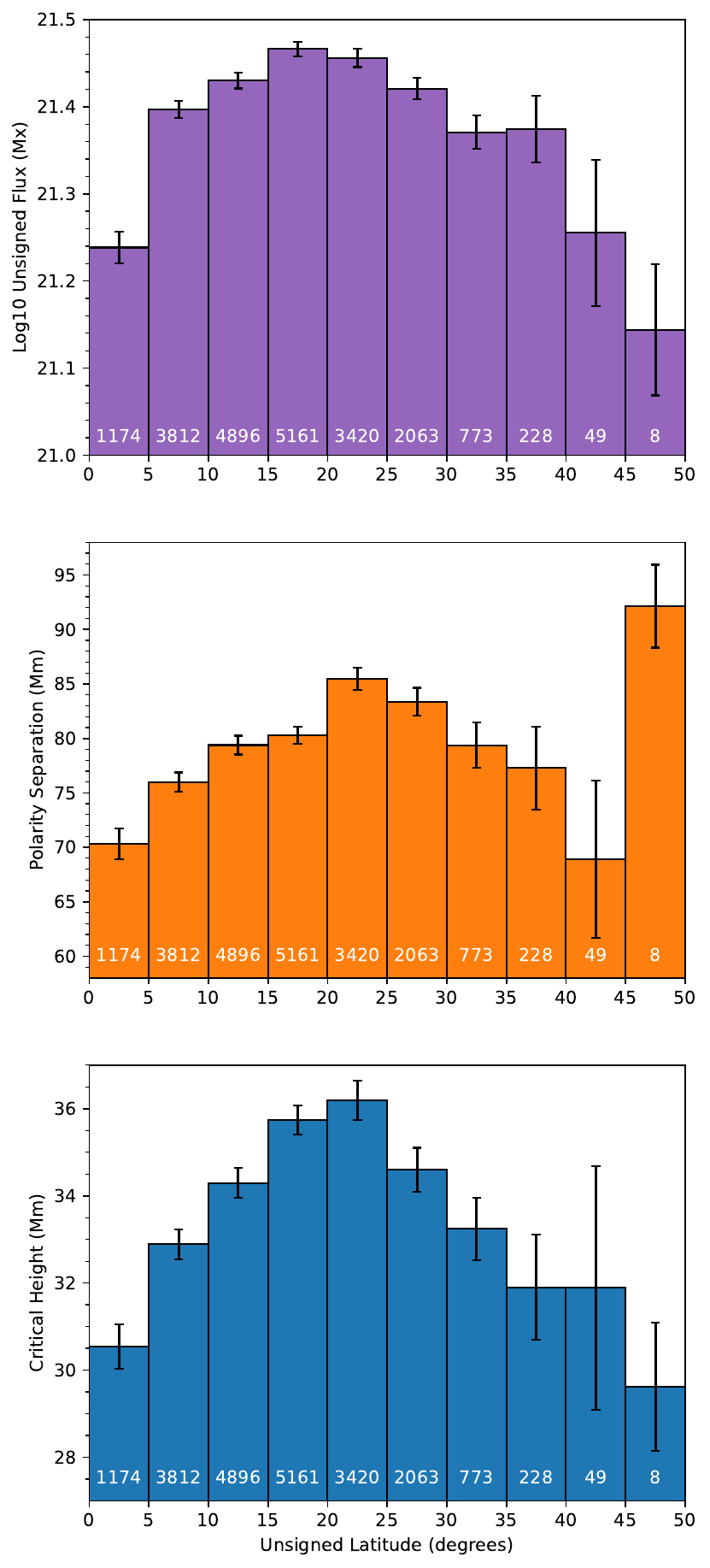}
    \caption{Median unsigned magnetic flux (top), polarity separation (middle), and critical height (bottom) of our 21584 magnetograms binned into unsigned latitudinal bands. The number of magnetograms in each latitudinal band is indicated by white text labels, and the standard error of the median is indicated with black errorbars.}
    \label{fig:latitude}
\end{figure}

As shown in Figure \ref{fig:latitude}, we bin the magnetic fluxes, polarity separations, and critical heights from our 21584 magnetograms into unsigned latitudinal bands of $5^{\circ}$ (\textit{i.e.} no distinction between north and south) and examine the median values in each band.
Firstly, there appear to be exceptionally large polarity separations found between $\pm 45\textrm{--}50^{\circ}$, but this is likely due to anomalously large values in the 8 magnetograms in our sample at these latitudes, which all come from the weak field region, NOAA AR 08175.

Ignoring this, we generally see the largest magnetic fluxes, polarity separations, and critical heights in the $\pm 15\textrm{--}25^{\circ}$ latitudinal range, with smaller values found further towards the equator and the poles.
This fits with the scenario where these parameters vary with the solar cycle, because the latitudes of our magnetic regions also vary with the solar cycle. As shown in the bottom panel of Figure \ref{fig:mgms}, magnetic regions typically lie at higher latitudes at the start of each solar cycle, and their latitude decreases over the next 11 years. 
Our results suggest that, during the activity minimum start of each 11-year cycle, magnetic regions emerge at high latitudes and have relatively small magnetic fluxes, polarity separations, and critical heights. By solar maximum, magnetic regions exist at lower latitudes ($\sim 20^{\circ}$) with stronger magnetic fluxes, larger polarity separations, and higher critical heights. Then, as the next sunspot minimum approaches, magnetic patches emerge closer to the equator, and their magnetic fluxes, polarity separations, and critical heights are smaller once again.

Around solar maximum, we do observe values in some magnetograms that are comparably low to those seen at solar minimum (see the lower ends of the standard deviation errorbar spreads in Figure \ref{fig:cycle_variations}). 
These low critical heights may occur in complex, multipolar active regions as suggested by \citealt{torok2007simulations}).
However, there are also many higher critical heights found each month around solar maximum.
In other words, there is a larger spread of high and low magnetic fluxes, polarity separations and critical heights around solar maximum, whereas there is only a small spread of relatively low values at solar minimum.
But more than just the difference in the range of values found from magnetograms at different points in the solar cycle, it is clear that the monthly mean values found in active regions are significantly greater than those seen at solar minimum. 

Despite the sunspot peak of solar cycle 23 being larger than in solar cycle 24, the mean monthly values of each parameter are not significantly different from one cycle to another. 
This suggests the variations seen in the magnetic flux, polarity separation and critical height throughout each cycle do not depend on the absolute number of sunspots (else, we would expect to see a difference in the flux, polarity separation and critical height values from one cycle to the next that depends on the peak sunspot number).
Instead, there appears to be an intrinsic evolution in the magnetic properties of the regions from solar minimum to solar maximum, irrespective of the magnitude of the sunspot number. 

Higher critical heights should make it harder for a flux rope to become torus unstable and erupt as a CME, as the flux rope would have to reach that higher instability height. 
But following this logic, the higher critical heights that we find in active regions around solar maximum appear to contradict the observation of more CMEs than at solar minimum.
The lower quantity of active regions could mean that fewer flux ropes form at solar minimum, explaining the relative lack of eruptions. But how are the many CMEs seen at solar maximum able to occur? 

Perhaps CMEs around solar maximum are driven by mechanisms other than the torus instability, such as magnetic reconnection.
Or perhaps many of the solar maximum CMEs come from the regions where we still observe low critical heights. After all, low critical heights are found in magnetically-complex active regions \citep{torok2007simulations}, which can produce many CMEs \citep{green2002prolific,gopalswamy2005extreme}. 
However, we may have excluded many of these complex CME-producing magnetic regions with low critical heights by selecting only ARPs that contain one or no NOAA active region.

We must also consider the limitations of the methodology we have used in this paper. When calculating the critical height for each magnetogram, we take the mean of the critical heights identified in all of the identified PIL pixels. 
However, where multiple distinct PILs exist, the critical height associated with one may be lower than the other. Our averaged critical height will be an overestimate for one PIL and an underestimate for the other, and a CME could erupt from either. We tried to minimize the occurrence of this scenario by excluding ARPs that contained multiple active regions and optimising our method of defining PILs, but the presence of complex active regions and new flux emerging into decayed ephemeral regions in our dataset means we still may mischaracterise the true critical height associated with a CME-producing ARP in some cases.

Still, for a statistically significant sample of 21584 magnetograms, our results show that the mean monthly critical height in active regions is higher at solar maximum than at solar minimum, and during many months around solar maximum, high critical heights are also observed in ephemeral regions.
Some pre-eruptive flux ropes may somehow be able to overcome these higher critical heights. 
Understanding the heights of flux ropes is just as important as knowing the critical height when determining whether the torus instability can cause an eruption.

The cancellation of magnetic flux in the photosphere and chromosphere can form low-altitude flux ropes \citep[\textit{e.g.}][]{yardley2018cancellation}. These flux ropes may have bald patch separatrix surface configurations \citep{titov1993bpss}, in which their underside is line-tied to the photosphere by high $\beta$ plasma. Similarly dense plasma can also manifest as an associated filament along dips in the helical magnetic field of a flux rope.

\citet{wang2010filaments} identified 9477 prominences (filaments observed at the solar limb) between 2007 and 2009. $99\%$ of these prominences were between $30^{\circ}$ and $60^{\circ}$ in latitude, and $82\%$ lay at a height of about $26\, \mathrm{Mm}$ above the photosphere. These prominences were observed around solar minimum, and this height of $26\, \mathrm{Mm}$ is comparable to the critical heights we observe around solar minimum in our study ($25 \pm 5\, \mathrm{Mm}$). However, such filaments would lie well below the typical active region critical heights we observe around solar maximum (roughly $35 \pm 10\, \mathrm{Mm}$).

On the other hand, quiescent filaments (long-lived and often appearing at high latitudes) tend to form or rise to much higher heights. \citet{cheng2020initiation} found the mean critical height at the onset of six quiescent filament eruptions to be $118.3 \pm 47.4\, \textrm{Mm}$.
Quiescent filaments tend to form during the decay phases of active regions which, after significant decay, may be classed as ephemeral regions), and the increased quantity of active regions around solar maximum means this can happen more frequently.
Indeed, the filament eruptions studied by \citet{cheng2020initiation} occurred between 2012 and 2014, demonstrating how quiescent filaments can form or rise to meet great critical heights around solar maximum periods.

We can also consider flux ropes that form in active regions without filaments. 
\citet{james2022evolution} studied 47 active region CMEs and found the mean critical height at eruption onset was $43 \pm 8\, \mathrm{Mm}$.
Furthermore, \citet{cheng2020initiation} found the mean critical height in ``hot channel'' eruptions was $58.0 \pm 33.6\, \textrm{Mm}$. These ``hot channels'' are signatures of heated plasma trapped in the magnetic field of active region flux ropes. 
These hot flux ropes can form via magnetic reconnection in the corona that is triggered by the motions of emerging magnetic flux \citep{james2017on-disc,james2020trigger}. 
They can form at heights of $\approx 100\, \mathrm{Mm}$ in the corona \citep{james2018model} with a hyperbolic flux tube configuration (HFT; \citealp{titov2002hft}), in which magnetic reconnection can occur in a current sheet around the flux rope, enabling it to rise and reach higher altitudes.

\citet{james2022evolution} found high CME rates during phases of increasing magnetic flux, even though the critical height was also increasing at the same time.
The result in the present study, that magnetic fluxes and critical heights are higher at solar maximum (when the most CMEs occur), seems to echo this scenario. 
We suggest that the effect magnetic flux emergence has on triggering a CME is greater than the extra stabilising effect provided by an associated rise in the critical height. 
Emerging magnetic flux could cause hot channel HFT flux ropes to form at relatively high coronal heights, and the photospheric motions associated with the flux emergence could inject magnetic energy into such a flux rope, causing it to inflate \citep{barnes1972inflation,torok2013inflation} and rise above an increased critical height. 
Additionally, newly-emerged magnetic flux could reconnect with --- and be built into --- a forming pre-eruptive flux rope, adding twist to the structure. This added twist could evolve a flux rope towards the threshold for the onset of the helical kink instability \citep{torok2005kink}, which could deform the flux rope axis and cause it to rise towards the critical torus instability height.

\section{Conclusions} \label{sec:conclusions}

Using a large sample of 21584 MDI and HMI magnetograms from 1996--2023 containing NOAA active regions and ephemeral regions, we find that the critical torus instability height varies with the solar sunspot cycle. 
The mean value of the critical height in months around solar cycle minima are found to be relatively low, and there are small spreads in the observed critical heights. In contrast, larger mean monthly critical heights occur around solar maximum, with larger spreads in the values that are seen each month.
We also see similar variations of the magnitude of unsigned magnetic flux and the separation between the centres of positive and negative magnetic polarities throughout solar cycles 23, 24 and 25, each also peaking around solar maximum and showing the lowest monthly values around solar minimum.

The critical height is strongly correlated with the separation between opposite magnetic polarities ($r=0.854$), and we fit a relationship of $h_\mathrm{c} = 0.5d + 7.35$ to the values found from our 21584 magnetograms. This is in close agreement with relations found in previous studies \citep{wang1989bipolar,baumgartner2018eruptive,james2022evolution}.
Secondly, we find that the logarithm of the critical height correlates well with the logarithm of magnetic flux ($r=0.738$), and thirdly, we provide an updated look at the connection between magnetic flux and polarity separation. Across our full dataset, we find the logarithm of magnetic flux correlates with the logarithm of polarity separation ($r=0.600$), and we fit the power law $\Phi \sim d^{1.23}$. This sits well between the results found in previous studies \citep{wang1989bipolar,tian2003bipolar}, and breaking the dataset down to separately examine active regions and ephemeral regions, we find relationships of $\Phi \sim d^{1.14}$ and $\Phi \sim d^{1.31}$, respectively.

In summary, all three parameters --- unsigned magnetic flux, the separation between opposite magnetic polarities, and the critical torus instability height --- correlate well with each other.
Therefore, we suggest the stronger magnetic fluxes found during periods of higher solar magnetic activity lead to larger separations (as emerging bipoles expand), and that the magnetic field strength of these larger-scale magnetic bipoles decays with height over larger length scales, resulting in higher critical heights.

Despite the higher critical heights found around solar maximum, more CMEs still occur then than at solar minimum. For the torus instability to play a role in driving these eruptions, the majority of CMEs must either originate from the regions that have the lowest critical heights, or the flux ropes that form around solar maximum must be able to overcome higher critical heights. 

\begin{acknowledgments}
We acknowledge the helpful discussions with K. D. Leka with respect to radialising MDI magnetic field observations.
A.W.J. was supported by a European Space Agency (ESA) Research Fellowship and acknowledges funding from the STFC Consolidated Grant ST/W001004/1.
Data courtesy of NASA/SDO and the HMI science team, as well as the SOHO/MDI consortium. SOHO is a project of international cooperation between ESA and NASA.
This research has made use of SunPy 4.1.3 \citep{sunpy_4.1.3}, an open-source and free community-developed solar data analysis Python package \citep{sunpy2020}.
\end{acknowledgments}

%

\vspace{5mm}
\facilities{SDO(HMI), SOHO(MDI)}








\bibliography{sample631}{}

\begin{thebibliography}{}
\expandafter\ifx\csname natexlab\endcsname\relax\def\natexlab#1{#1}\fi
\providecommand{\url}[1]{\href{#1}{#1}}
\providecommand{\dodoi}[1]{doi:~\href{http://doi.org/#1}{\nolinkurl{#1}}}
\providecommand{\doeprint}[1]{\href{http://ascl.net/#1}{\nolinkurl{http://ascl.net/#1}}}
\providecommand{\doarXiv}[1]{\href{https://arxiv.org/abs/#1}{\nolinkurl{https://arxiv.org/abs/#1}}}

\bibitem[{{Alissandrakis}(1981)}]{alissandrakis1981field}
{Alissandrakis}, C.~E. 1981, \aap, 100, 197

\bibitem[{{Barnes} \& {Sturrock}(1972)}]{barnes1972inflation}
{Barnes}, C.~W., \& {Sturrock}, P.~A. 1972, \apj, 174, 659,
  \dodoi{10.1086/151527}

\bibitem[{{Bateman}(1978)}]{bateman1978instabilities}
{Bateman}, G. 1978, {MHD instabilities} (MIT Press)

\bibitem[{{Baumgartner} {et~al.}(2018){Baumgartner}, {Thalmann}, \&
  {Veronig}}]{baumgartner2018eruptive}
{Baumgartner}, C., {Thalmann}, J.~K., \& {Veronig}, A.~M. 2018, \apj, 853, 105,
  \dodoi{10.3847/1538-4357/aaa243}

\bibitem[{{Bhowmik} \& {Nandy}(2018)}]{bhowmik2018prediction}
{Bhowmik}, P., \& {Nandy}, D. 2018, Nature Communications, 9, 5209,
  \dodoi{10.1038/s41467-018-07690-0}

\bibitem[{{Bobra} {et~al.}(2014){Bobra}, {Sun}, {Hoeksema}, {Turmon}, {Liu},
  {Hayashi}, {Barnes}, \& {Leka}}]{bobra2014sharps}
{Bobra}, M.~G., {Sun}, X., {Hoeksema}, J.~T., {et~al.} 2014, \solphys, 289,
  3549, \dodoi{10.1007/s11207-014-0529-3}

\bibitem[{{Bobra} {et~al.}(2021){Bobra}, {Wright}, {Sun}, \&
  {Turmon}}]{bobra2021smarps}
{Bobra}, M.~G., {Wright}, P.~J., {Sun}, X., \& {Turmon}, M.~J. 2021, \apjs,
  256, 26, \dodoi{10.3847/1538-4365/ac1f1d}

\bibitem[{{Borrero} {et~al.}(2011){Borrero}, {Tomczyk}, {Kubo},
  {Socas-Navarro}, {Schou}, {Couvidat}, \& {Bogart}}]{borrero2011VFISV}
{Borrero}, J.~M., {Tomczyk}, S., {Kubo}, M., {et~al.} 2011, \solphys, 273, 267,
  \dodoi{10.1007/s11207-010-9515-6}

\bibitem[{{Cheng} {et~al.}(2020){Cheng}, {Zhang}, {Kliem}, {T{\"o}r{\"o}k},
  {Xing}, {Zhou}, {Inhester}, \& {Ding}}]{cheng2020initiation}
{Cheng}, X., {Zhang}, J., {Kliem}, B., {et~al.} 2020, \apj, 894, 85,
  \dodoi{10.3847/1538-4357/ab886a}

\bibitem[{{D{\'e}moulin} \& {Aulanier}(2010)}]{demoulin2010criteria}
{D{\'e}moulin}, P., \& {Aulanier}, G. 2010, \apj, 718, 1388,
  \dodoi{10.1088/0004-637X/718/2/1388}

\bibitem[{{Domingo} {et~al.}(1995){Domingo}, {Fleck}, \&
  {Poland}}]{domingo1995soho}
{Domingo}, V., {Fleck}, B., \& {Poland}, A.~I. 1995, \solphys, 162, 1,
  \dodoi{10.1007/BF00733425}

\bibitem[{{Fan} \& {Gibson}(2007)}]{fan2007onset}
{Fan}, Y., \& {Gibson}, S.~E. 2007, \apj, 668, 1232, \dodoi{10.1086/521335}

\bibitem[{{Gopalswamy} {et~al.}(2005){Gopalswamy}, {Yashiro}, {Liu},
  {Michalek}, {Vourlidas}, {Kaiser}, \& {Howard}}]{gopalswamy2005extreme}
{Gopalswamy}, N., {Yashiro}, S., {Liu}, Y., {et~al.} 2005, Journal of
  Geophysical Research (Space Physics), 110, A09S15,
  \dodoi{10.1029/2004JA010958}

\bibitem[{{Gosain} {et~al.}(2016){Gosain}, {Filippov}, {Ajor Maurya}, \&
  {Chandra}}]{gosain2016interrupted}
{Gosain}, S., {Filippov}, B., {Ajor Maurya}, R., \& {Chandra}, R. 2016, \apj,
  821, 85, \dodoi{10.3847/0004-637X/821/2/85}

\bibitem[{{Green} {et~al.}(2002){Green}, {L{\'o}pez fuentes}, {Mandrini},
  {D{\'e}moulin}, {Van Driel-Gesztelyi}, \& {Culhane}}]{green2002prolific}
{Green}, L.~M., {L{\'o}pez fuentes}, M.~C., {Mandrini}, C.~H., {et~al.} 2002,
  \solphys, 208, 43, \dodoi{10.1023/A:1019658520033}

\bibitem[{{Guo} {et~al.}(2010{\natexlab{a}}){Guo}, {Zhang}, {Chumak}, \&
  {Lin}}]{guo2010complexity}
{Guo}, J., {Zhang}, H.~Q., {Chumak}, O.~V., \& {Lin}, J.~B. 2010{\natexlab{a}},
  \mnras, 405, 111, \dodoi{10.1111/j.1365-2966.2010.16465.x}

\bibitem[{{Guo} {et~al.}(2010{\natexlab{b}}){Guo}, {Ding}, {Schmieder}, {Li},
  {T{\"o}r{\"o}k}, \& {Wiegelmann}}]{guo2010confined}
{Guo}, Y., {Ding}, M.~D., {Schmieder}, B., {et~al.} 2010{\natexlab{b}}, \apjl,
  725, L38, \dodoi{10.1088/2041-8205/725/1/L38}

\bibitem[{{James} {et~al.}(2020){James}, {Green}, {van Driel-Gesztelyi}, \&
  {Valori}}]{james2020trigger}
{James}, A.~W., {Green}, L.~M., {van Driel-Gesztelyi}, L., \& {Valori}, G.
  2020, \aap, 644, A137, \dodoi{10.1051/0004-6361/202038781}

\bibitem[{{James} {et~al.}(2018){James}, {Valori}, {Green}, {Liu}, {Cheung},
  {Guo}, \& {van Driel-Gesztelyi}}]{james2018model}
{James}, A.~W., {Valori}, G., {Green}, L.~M., {et~al.} 2018, \apjl, 855, L16,
  \dodoi{10.3847/2041-8213/aab15d}

\bibitem[{{James} {et~al.}(2022){James}, {Williams}, \&
  {O'Kane}}]{james2022evolution}
{James}, A.~W., {Williams}, D.~R., \& {O'Kane}, J. 2022, \aap, 665, A37,
  \dodoi{10.1051/0004-6361/202142910}

\bibitem[{{James} {et~al.}(2017){James}, {Green}, {Palmerio}, {Valori}, {Reid},
  {Baker}, {Brooks}, {van Driel-Gesztelyi}, \& {Kilpua}}]{james2017on-disc}
{James}, A.~W., {Green}, L.~M., {Palmerio}, E., {et~al.} 2017, \solphys, 292,
  71, \dodoi{10.1007/s11207-017-1093-4}

\bibitem[{{Kliem} \& {T{\"o}r{\"o}k}(2006)}]{kliem2006torus}
{Kliem}, B., \& {T{\"o}r{\"o}k}, T. 2006, \prl, 96, 255002,
  \dodoi{10.1103/PhysRevLett.96.255002}

\bibitem[{{Leka} {et~al.}(2017){Leka}, {Barnes}, \& {Wagner}}]{leka2017radial}
{Leka}, K.~D., {Barnes}, G., \& {Wagner}, E.~L. 2017, \solphys, 292, 36,
  \dodoi{10.1007/s11207-017-1057-8}

\bibitem[{{Li} {et~al.}(2020){Li}, {Hou}, {Yang}, {Zhang}, {Liu}, \&
  {Veronig}}]{li2020fluxcritheight}
{Li}, T., {Hou}, Y., {Yang}, S., {et~al.} 2020, \apj, 900, 128,
  \dodoi{10.3847/1538-4357/aba6ef}

\bibitem[{{Liu} {et~al.}(2018){Liu}, {Wang}, {Zhou}, {Dissauer}, {Temmer}, \&
  {Cui}}]{liu2018failed}
{Liu}, L., {Wang}, Y., {Zhou}, Z., {et~al.} 2018, \apj, 858, 121,
  \dodoi{10.3847/1538-4357/aabba2}

\bibitem[{{Liu}(2008)}]{liu2008instabilites}
{Liu}, Y. 2008, \apjl, 679, L151, \dodoi{10.1086/589282}

\bibitem[{{Luo} \& {Liu}(2022)}]{luo2022saddles}
{Luo}, R., \& {Liu}, R. 2022, arXiv e-prints, Accepted to \apj,
  arXiv:2203.03913.
\newblock \doarXiv{2203.03913}

\bibitem[{{McIntosh} {et~al.}(2023){McIntosh}, {Leamon}, \&
  {Egeland}}]{mcintosh2023terminator}
{McIntosh}, S.~W., {Leamon}, R.~J., \& {Egeland}, R. 2023, Frontiers in
  Astronomy and Space Sciences, 10, 16, \dodoi{10.3389/fspas.2023.1050523}

\bibitem[{Mumford {et~al.}(2023)Mumford, Freij, Stansby, Christe, Ireland,
  Mayer, Shih, Hughitt, Ryan, Liedtke, Hayes, Pérez-Suárez, I., Barnes,
  Chakraborty, Inglis, Pattnaik, Sipőcz, MacBride, Sharma, Leonard, Hewett,
  Hamilton, Manhas, Panda, Earnshaw, Choudhary, Kumar, Singh, Chanda, Haque,
  Kirk, Mueller, Konge, Srivastava, Wentzel-Long, Jain, Bennett, Baruah,
  Arbolante, Charlton, Maloney, Mishra, Paul, Verma, Chorley, Chouhan,
  Himanshu, Mason, Zivadinovic, Modi, Sharma, Naman9639, Bobra, Rozo, Manley,
  Ivashkiv, Laitinen, Chatterjee, von Forstner, Bazán, Stern, Gieseler, Evans,
  Jain, Malocha, Ghosh, Airmansmith97, Stańczak, Singh, Visscher, Verma,
  SophieLemos, Agrawal, Alam, Buddhika, Pathak, Rideout, Sharma, Park, Bates,
  Wilson, Shukla, Giger, Mishra, Sharma, Goel, Taylor, Cetusic, Reiter, Jacob,
  Inchaurrandieta, Dacie, Dubey, Eigenbrot, Bray, Surve, Zahniy, Sidhu,
  Meszaros, Parkhi, Russell, Bose, Pandey, Price-Whelan, J, Chicrala, Ankit,
  Guennou, D'Avella, Williams, Verma, Ballew, Agrawal, Murphy, Lodha,
  Robitaille, Augspurger, Krishan, honey, neerajkulk, Bhope, Gaba, Hill,
  Mampaey, Wiedemann, Molina, Briseno, Keşkek, Habib, Letts, Singaravelan,
  Ranjan, Altunian, Streicher, Gomillion, Agarwal, Kothari, Nomiya,
  mridulpandey, Stevens, B, Bahuleyan, Kaszynski, W, Mehrotra, Tang, Sinha,
  Smith, Kustov, Stone, Bard, Arias, Tollerud, Dover, Verstringe, Kumar,
  Mathur, Babuschkin, Calixto, Wimbish, Qing, Buitrago-Casas, Krishna,
  Chaudhari, Hiware, Ghosh, Lyes, Mangaonkar, Cheung, Mendero, Dedhia,
  Schoentgen, Shahdadpuri, Srinivasan, Gyenge, Mekala, Das, Mishra, Sharma,
  Srikanth, Jain, Kannojia, Yadav, Paul, Wilkinson, Caswell, Braccia, Pereira,
  Gates, Dang, Bankar, Jamieson, Agrawal, platipo, resakra, tal66, yasintoda,
  Attie, \& Murray}]{sunpy_4.1.3}
Mumford, S.~J., Freij, N., Stansby, D., {et~al.} 2023, SunPy, v4.1.3,  Zenodo,
  \dodoi{10.5281/zenodo.7641693}

\bibitem[{NOAA(2023)}]{noaa2023update}
NOAA. 2023, \textit{NOAA forecasts quicker, stronger peak of solar activity}

\bibitem[{{Pesnell} {et~al.}(2012){Pesnell}, {Thompson}, \&
  {Chamberlin}}]{pesnell2012SDO}
{Pesnell}, W.~D., {Thompson}, B.~J., \& {Chamberlin}, P.~C. 2012, \solphys,
  275, 3, \dodoi{10.1007/s11207-011-9841-3}

\bibitem[{{Scherrer} {et~al.}(1995){Scherrer}, {Bogart}, {Bush}, {Hoeksema},
  {Kosovichev}, {Schou}, {Rosenberg}, {Springer}, {Tarbell}, {Title},
  {Wolfson}, {Zayer}, \& {MDI Engineering Team}}]{scherrer1995mdi}
{Scherrer}, P.~H., {Bogart}, R.~S., {Bush}, R.~I., {et~al.} 1995, \solphys,
  162, 129, \dodoi{10.1007/BF00733429}

\bibitem[{{Scherrer} {et~al.}(2012){Scherrer}, {Schou}, {Bush}, {Kosovichev},
  {Bogart}, {Hoeksema}, {Liu}, {Duvall}, {Zhao}, {Title}, {Schrijver},
  {Tarbell}, \& {Tomczyk}}]{scherrer2012hmi}
{Scherrer}, P.~H., {Schou}, J., {Bush}, R.~I., {et~al.} 2012, \solphys, 275,
  207, \dodoi{10.1007/s11207-011-9834-2}

\bibitem[{{Schrijver}(2007)}]{schrijver2007R}
{Schrijver}, C.~J. 2007, \apjl, 655, L117, \dodoi{10.1086/511857}

\bibitem[{{Sun} {et~al.}(2022){Sun}, {T{\"o}r{\"o}k}, \&
  {DeRosa}}]{sun2022torus}
{Sun}, X., {T{\"o}r{\"o}k}, T., \& {DeRosa}, M.~L. 2022, \mnras, 509, 5075,
  \dodoi{10.1093/mnras/stab3249}

\bibitem[{{SunPy Project} {et~al.}(2020){SunPy Project}, {Barnes}, {Bobra},
  {Christe}, {Freij}, {Hayes}, {Ireland }, {Mumford}, {Perez-Suarez}, {Ryan},
  {Shih}, {Chanda}, {Glogowski}, {Hewett}, {Hughitt}, {Hill}, {Hiware},
  {Inglis}, {Kirk}, {Konge}, {Mason}, {Maloney}, {Murray}, {Panda}, {Park},
  {Pereira}, {Reardon}, {Savage}, {Sip{\H{o}}cz}, {Stansby}, {Jain}, {Taylor},
  {Yadav}, {Rajul}, \& {Dang}}]{sunpy2020}
{SunPy Project}, {Barnes}, W.~T., {Bobra}, M.~G., {et~al.} 2020, \apj, 890, 68,
  \dodoi{10.3847/1538-4357/ab4f7a}

\bibitem[{{Tian} {et~al.}(2003){Tian}, {Liu}, \& {Wang}}]{tian2003bipolar}
{Tian}, L., {Liu}, Y., \& {Wang}, H. 2003, \solphys, 215, 281,
  \dodoi{10.1023/A:1025686305225}

\bibitem[{{Titov} {et~al.}(2002){Titov}, {Hornig}, \&
  {D{\'e}moulin}}]{titov2002hft}
{Titov}, V.~S., {Hornig}, G., \& {D{\'e}moulin}, P. 2002, Journal of
  Geophysical Research (Space Physics), 107, 1164, \dodoi{10.1029/2001JA000278}

\bibitem[{{Titov} {et~al.}(1993){Titov}, {Priest}, \&
  {Demoulin}}]{titov1993bpss}
{Titov}, V.~S., {Priest}, E.~R., \& {Demoulin}, P. 1993, \aap, 276, 564

\bibitem[{{T{\"o}r{\"o}k} \& {Kliem}(2005)}]{torok2005kink}
{T{\"o}r{\"o}k}, T., \& {Kliem}, B. 2005, \apjl, 630, L97,
  \dodoi{10.1086/462412}

\bibitem[{{T{\"o}r{\"o}k} \& {Kliem}(2007)}]{torok2007simulations}
---. 2007, Astronomische Nachrichten, 328, 743, \dodoi{10.1002/asna.200710795}

\bibitem[{{T{\"o}r{\"o}k} {et~al.}(2013){T{\"o}r{\"o}k}, {Temmer}, {Valori},
  {Veronig}, {van Driel-Gesztelyi}, \& {Vr{\v{s}}nak}}]{torok2013inflation}
{T{\"o}r{\"o}k}, T., {Temmer}, M., {Valori}, G., {et~al.} 2013, \solphys, 286,
  453, \dodoi{10.1007/s11207-013-0269-9}

\bibitem[{{Wang} {et~al.}(2017){Wang}, {Liu}, {Wang}, {Liu}, {Chen}, {Liu},
  {Zhou}, \& {Zhang}}]{wang2017critical}
{Wang}, D., {Liu}, R., {Wang}, Y., {et~al.} 2017, \apjl, 843, L9,
  \dodoi{10.3847/2041-8213/aa79f0}

\bibitem[{{Wang} {et~al.}(2010){Wang}, {Cao}, {Chen}, {Zhang}, {Yu}, {Zheng},
  {Shen}, {Zhang}, \& {Wang}}]{wang2010filaments}
{Wang}, Y., {Cao}, H., {Chen}, J., {et~al.} 2010, \apj, 717, 973,
  \dodoi{10.1088/0004-637X/717/2/973}

\bibitem[{{Wang} \& {Sheeley}(1989)}]{wang1989bipolar}
{Wang}, Y.~M., \& {Sheeley}, N.~R., J. 1989, \solphys, 124, 81,
  \dodoi{10.1007/BF00146521}

\bibitem[{{Webb} \& {Howard}(1994)}]{webb1994cycle}
{Webb}, D.~F., \& {Howard}, R.~A. 1994, \jgr, 99, 4201,
  \dodoi{10.1029/93JA02742}

\bibitem[{{Webb} \& {Howard}(2012)}]{webb2012cmes}
{Webb}, D.~F., \& {Howard}, T.~A. 2012, Living Reviews in Solar Physics, 9, 3,
  \dodoi{10.12942/lrsp-2012-3}

\bibitem[{{Yardley} {et~al.}(2018){Yardley}, {Green}, {van Driel-Gesztelyi},
  {Williams}, \& {Mackay}}]{yardley2018cancellation}
{Yardley}, S.~L., {Green}, L.~M., {van Driel-Gesztelyi}, L., {Williams}, D.~R.,
  \& {Mackay}, D.~H. 2018, \apj, 866, 8, \dodoi{10.3847/1538-4357/aade4a}

\bibitem[{{Zuccarello} {et~al.}(2015){Zuccarello}, {Aulanier}, \&
  {Gilchrist}}]{zuccarello2015critical}
{Zuccarello}, F.~P., {Aulanier}, G., \& {Gilchrist}, S.~A. 2015, \apj, 814,
  126, \dodoi{10.1088/0004-637X/814/2/126}

\end{thebibliography}
\bibliographystyle{aasjournal}



\end{document}